\newif\ifonecolumn
\onecolumnfalse
\newif\iftwocolumn

\ifonecolumn
    \twocolumnfalse
\else
    \twocolumntrue
    \newcommand\spaceunderfig{-15pt}
\fi

\ifonecolumn
    \documentclass[12pt, draftclsnofoot, onecolumn]{IEEEtran}

    \usepackage[font=footnotesize]{subcaption}
    \usepackage[export]{adjustbox}

    \usepackage{microtype} 	% Kirjain- ja sanavälin muokkaus paremmaksi (+ rivitys)
    \usepackage{relsize}	% Fontin skaalaus

    \newcommand\spaceunderfig{-40pt}

\else
\documentclass[twocolumn]{IEEEtran}
\fi

\newcommand\speciallinespacing{1.00}

    \usepackage[export]{adjustbox}

    \usepackage{microtype} 	% Kirjain- ja sanavälin muokkaus paremmaksi (+ rivitys)
    \usepackage{relsize}	% Fontin skaalaus

\usepackage{graphicx}
\usepackage{cite}
\usepackage{multirow, array}
\usepackage[cmex10]{amsmath}
\usepackage{amssymb}
\usepackage{cases}
\usepackage[font=footnotesize]{caption}
\usepackage{siunitx}
\usepackage{xcolor}
\usepackage{placeins}
\usepackage{xspace}
\usepackage{textcomp}
\usepackage{import}
\usepackage[update,prepend]{epstopdf}
\usepackage{microtype}
\usepackage[nonumberlist]{glossaries}
\usepackage{bm}
\usepackage[super]{nth}
\usepackage{relsize}
\usepackage{url}
\usepackage{algorithm, algpseudocode}
\usepackage[breaklinks=true]{hyperref}

\newacronym{5G}{5G}{fifth generation}
\newacronym{ACF}{ACF}{autocorrelation function}
\newacronym{AR}{AR}{auto-regressive}
\newacronym{AN}{AN}{access node}
\newacronym{AoA}{AoA}{angle of arrival}
\newacronym{AWGN}{AWGN}{additive white Gaussian noise}
\newacronym{BS}{BS}{base station}
\newacronym{BIC}{BIC}{Bayesian information criterion}
\newacronym{BPSK}{BPSK}{binary phase shift keying}
\newacronym{CL}{CL}{centroid localization}
\newacronym{CR}{CR}{cognitive radio}
\newacronym{CRB}{CRB}{Cramer-Rao bound}
\newacronym{CSI}{CSI}{channel state information}
\newacronym{CSIT}{CSIT}{channel state information at transmitter}
\newacronym{CA}{CA}{constant acceleration}
\newacronym{CT}{CT}{constant turn}
\newacronym{CV}{CV}{constant velocity}
\newacronym{D2D}{D2D}{device-to-device}
\newacronym{DBS}{DBS}{different beamwidth sectors}
\newacronym{DCAA}{DCAA}{digitally controlled antenna array}
\newacronym{DFU}{DFU}{{}\acrshort{DoA} fusion}
\newacronym{DL}{DL}{downlink}
\newacronym[firstplural=directions of arrival (DoAs)]{DoA}{DoA}{direction of arrival}
\newacronym{DS-CDMA}{DC-CDMA}{direct sequence code division multiple access}
\newacronym{E2E}{E2E}{end-to-end}
\newacronym{E911}{E911}{enhanced 911}
\newacronym{EADF}{EADF}{effective aperture distribution function}
\newacronym{EBS}{EBS}{equal beamwidth sectors}
\newacronym{EKF}{EKF}{extended Kalman filter}
\newacronym{ESA}{ESA}{equal sector antenna}
\newacronym{ESPAR}{ESPAR}{electronically steerable parasitic array radiator}
\newacronym{EOTD}{E-OTD}{enhanced observed time difference}
\newacronym{EW}{EW}{equal weighting}
\newacronym{FCC}{FCC}{federal communications commission}
\newacronym{FFT}{FFT}{fast Fourier transform}
\newacronym[firstplural=Fisher information matrices (FIM)]{FIM}{FIM}{Fisher information matrix}
\newacronym{GMM}{GMM}{Gaussian mixture model}
\newacronym{GNSS}{GNSS}{global navigation satellite system}
\newacronym{GPS}{GPS}{global positioning system}
\newacronym{GSM}{GSM}{global system for mobile communications}
\newacronym{ICI}{ICI}{inter-carrier-interference}
\newacronym{ICMP}{ICMP}{internet control message protocol}
\newacronym{IoT}{IoT}{Internet-of-Things}
\newacronym{ISD}{ISD}{inter-site distance}
\newacronym{ISI}{ISI}{inter-symbol-interference}
\newacronym{ITS}{ITS}{intelligent traffic system}
\newacronym{ITU}{ITU}{international telecommunication union}
\newacronym{KF}{KF}{Kalman filter}
\newacronym{LMMSE}{LMMSE}{linear minimum mean-square error}
\newacronym{LoS}{LoS}{line-of-sight}
\newacronym{LS}{LS}{least squares}
\newacronym{LTE}{LTE}{long term evolution}
\newacronym{LWA}{LWA}{leaky-wave antenna}
\newacronym{MAP}{MAP}{maximum a posteriori} 
\newacronym{METIS}{METIS}{Mobile and wireless communications enablers for the twenty-twenty information society}
\newacronym{MGSCM}{MGSCM}{METIS geometry-based stochastic channel model}
\newacronym{MIMO}{MIMO}{multiple-input multiple-output}
\newacronym{MSE}{MSE}{mean-squared error}
\newacronym{ML}{ML}{maximum likelihood}
\newacronym{MLE}{MLE}{maximum likelihood estimator}
\newacronym{MMSE}{MMSE}{minimum mean-square error}
\newacronym{MU-MIMO}{MU-MIMO}{multiuser multiple-input-multiple-output}
\newacronym{MVUE}{MVUE}{minimum variance unbiased estimator}
\newacronym{NLoS}{NLoS}{non-line-of-sight}
\newacronym{ppm}{ppm}{parts per million}
\newacronym{OFDM}{OFDM}{orthogonal frequency-division multiplexing}
\newacronym{OFDMA}{OFDMA}{orthogonal frequency-division multiple access}
\newacronym{ON}{ON}{observing node}
\newacronym{OTDoA}{OTDoA}{observed time difference of arrival}
\newacronym{PDF}{PDF}{probability density function}
\newacronym{PDoA}{PDoA}{phase difference of arrival}
\newacronym{RToA}{RToA}{round-trip time of arrival}
\newacronym{PPMCC}{PPMCC}{Pearson product-moment correlation coefficient}
\newacronym{PU}{PU}{primary user}
\newacronym{PW}{PW}{power weighting}
\newacronym{RMSE}{RMSE}{root-mean-squared error}
\newacronym{RRMSE}{RRMSE}{relative root-mean-squared error}
\newacronym{RS}{RS}{reference signal}
\newacronym{RSS}{RSS}{received signal strength}
\newacronym{RRM}{RRM}{radio resource management}
\newacronym{RF}{RF}{radio frequency}
\newacronym{RFI}{RFI}{radio frequency interference}
\newacronym{SBS}{SBS}{switched-beam system}
\newacronym{SDE}{SDE}{sector-pair \acrshort{DoA} estimation}
\newacronym{SIMO}{SIMO}{single-input-multiple-output}
\newacronym{SINR}{SINR}{signal-to-interference-plus-noise ratio}
\newacronym{SLS}{SLS}{simplified least squares}
\newacronym{SNR}{SNR}{signal-to-noise ratio}
\newacronym{SSP}{SSP}{side-sector suppression}
\newacronym{SSL}{SSL}{sector selection}
\newacronym{STD}{STD}{standard deviation}
\newacronym{SU}{SU}{secondary user}
\newacronym{TCP}{TCP}{transmission control protocol}
\newacronym{TDD}{TDD}{time division duplex}
\newacronym{TDoA}{TDoA}{time difference of arrival}
\newacronym{TN}{TN}{target node}
\newacronym[firstplural=times of arrival (ToA)]{ToA}{ToA}{time of arrival}
\newacronym{ToF}{ToF}{time of flight}
\newacronym{TSLS}{TSLS}{three-stage SLS}
\newacronym{TTI}{TTI}{transmit time interval}
\newacronym{UAV}{UAV}{unmanned aerial vehicle}
\newacronym{UDN}{UDN}{ultra-dense network}
\newacronym{UKF}{UKF}{unscented Kalman filter}
\newacronym{UL}{UL}{uplink}
\newacronym{UTDoA}{U-TDoA}{uplink-time difference of arrival}
\newacronym{UTD}{UTD}{the uniform theory of diffraction}
\newacronym{UMi}{UMi}{urban micro}
\newacronym{UN}{UN}{user node}
\newacronym{VW}{VW}{variance weighting}
\newacronym{WCL}{WCL}{weighted centroid localization}
\newacronym{WLAN}{WLAN}{wireless local area network}
\newacronym{wrt}{wrt}{with respect to}
\newacronym{WSS}{WSS}{wide sense stationary}
\newif\ifpublicversion
\publicversionfalse
\newif\ifsmallfigures
\smallfiguresfalse

\DeclareMathOperator{\blkdiag}{blkdiag}

\DeclareMathOperator{\diag}{diag}

\renewcommand{\vec}[1]{\ensuremath{\boldsymbol{\mathbf{#1}}}}
\newcommand{\mat}[1]{\ensuremath{\boldsymbol{\mathbf{#1}}}}
\newcommand*{\eye}[1][]{%
\ifthenelse{\equal{#1}{}}{\ensuremath{\mat{I}}}{\ensuremath{\mat{I}_{#1}}}%
}
\newcommand*{\zeros}[1][]{%
\ifthenelse{\equal{#1}{}}{\ensuremath{\mat{0}}}{\ensuremath{\mat{0}_{#1}}}%
}
\newcommand*{\ones}[1][]{%
\ifthenelse{\equal{#1}{}}{\ensuremath{\mat{1}}}{\ensuremath{\mat{1}_{#1}}}%
}
\newcommand*{\transpose}{\ensuremath{^ \textrm{T}}}

\usepackage{pifont}
\newcolumntype{C}[1]{>{\centering\let\newline\\\arraybackslash\hspace{0pt}}m{#1}}
\newcolumntype{L}[1]{>{\raggedright\let\newline\\\arraybackslash\hspace{0pt}}m{#1}}

\newcommand{\doa}{\ensuremath{\gls{doa}}}

\newcommand{\sdoa}[1][]{%
\ifthenelse{\equal{#1}{}}{\ensuremath{\ensuremath{\sigma_{\doa}}}\xspace}{\ensuremath{\sigma_{\doa, #1}}\xspace}%
}

\ifpublicversion
	
	\newcommand*{\REFC}[1][REF]{\xspace}
\else
	
	\newcommand*{\REFC}[1][REF]{~{\color{blue}[#1]}\xspace}
\fi

\ifpublicversion
	\newcommand*{\SCOM}[1]{}

	\newcommand*{\ACOM}[1]{}
\else
	\newcommand*{\SCOM}[1]{{\color{blue} \it #1}\xspace}

	\newcommand*{\ACOM}[1]{{\color{red} \it #1}\xspace}
\fi

\newcommand{\scalethanks}[1]{

{\linespread{\speciallinespacing}
\let\thefootnote\relax\footnotetext{
\footnotesize
}}}

\newcommand{\TOA}{\gls{ToA}\xspace}
\newcommand{\TOAs}{\glspl{ToA}\xspace}
\newcommand{\DOA}{\gls{DoA}\xspace}
\newcommand{\DOAs}{\glspl{DoA}\xspace}

\newcommand{\DOATOAEKF}{\DOA/\TOA \gls{EKF}\xspace}
\newcommand{\DOATOAUKF}{\DOA/\TOA \gls{UKF}\xspace}
\newcommand{\DOATOAUKFs}{\DOA/\TOA \glspl{UKF}\xspace}
\newcommand{\LOSAN}{\gls{LoS}-\gls{AN}\xspace}
\newcommand{\LOSANs}{\gls{LoS}-\glspl{AN}\xspace}

\newcommand{\POSSYNCUKF}{\DOA/\TOA Pos\&Sync \gls{UKF}\xspace}
\newcommand{\POSCLOCKUKF}{\DOA/\TOA Pos\&Clock \gls{UKF}\xspace}

\newlength\myindent
\begin{document}

\bstctlcite{IEEEexample:BSTcontrol}

% Font scaling
%\ifonecolumn
    \relscale{0.95}
%\fi

%\abovedisplayskip=6pt
%\belowdisplayskip=6pt
\pagenumbering{gobble}

%\title{Joint User Node Positioning and Clock Synchronization in 5G Ultra-Dense Networks}

%\title{Joint 3D Positioning and Network Synchronization in {5G} Networks Using Unscented Kalman Filter}
%\title{Joint 3D Positioning and Network Synchronization in Future {5G} Networks Using UKF}
\title{Joint 3D Positioning and Network Synchronization in {5G} Ultra-Dense Networks Using UKF and EKF}

\author{
\IEEEauthorblockN{Mike Koivisto\IEEEauthorrefmark{1}, M\'ario Costa\IEEEauthorrefmark{2}, Aki Hakkarainen\IEEEauthorrefmark{1}, Kari Lepp\"anen\IEEEauthorrefmark{2}, and Mikko Valkama\IEEEauthorrefmark{1}}
%\IEEEauthorblockN{Janis Werner\IEEEauthorrefmark{1}, Mario Costa\IEEEauthorrefmark{2}, Aki Hakkarainen\IEEEauthorrefmark{1}, Kari Leppanen\IEEEauthorrefmark{2} and Mikko Valkama\IEEEauthorrefmark{1}}

\IEEEauthorblockA{\IEEEauthorrefmark{1} Department of Electronics and Communications Engineering, Tampere University of Technology, Finland\\
Emails: \{mike.koivisto, aki.hakkarainen, mikko.e.valkama\}@tut.fi}

\IEEEauthorblockA{\IEEEauthorrefmark{2} Huawei Technologies Oy (Finland) Co., Ltd, Finland R\&D Center\\
Emails: \{mariocosta, kari.leppanen\}@huawei.com}
\vspace{-5pt}
}%

%\thanks{M.~Koivisto, J.~Werner, J.~Talvitie and M.~Valkama are with the Department of Electronics and Communications Engineering, Tampere University of Technology, FI-33101 Tampere, Finland (email: \{mike.koivisto, janis.werner, jukka.talvitie, mikko.e.valkama\}@tut.fi).}
%\thanks{M.~Costa, K.~Heiska, and K.~Lepp\"anen are with Huawei Technologies Oy (Finland) Co., Ltd, Helsinki 00180, Finland (email: \{mariocosta, kari.heiska, kari.leppanen\}@huawei.com).}
%\thanks{V.~Koivunen is with the Department of Signal Processing and Acoustics, Aalto University, FI-02150 Espoo, Finland (email: visa.koivunen@aalto.fi).}
%\thanks{This work was supported by the Finnish Funding Agency for Technology and Innovation (Tekes), under the projects \textquotedblleft{5G Networks and Device Positioning\textquotedblright}, and \textquotedblleft{Future Small-Cell Networks using Reconfigurable Antennas\textquotedblright}.}
%\thanks{Preliminary work addressing a limited subset of initial results was presented at IEEE Global Communications Conference (GLOBECOM), San Diego, CA, USA, December 2015~\cite{werner_joint_2015}.}
%\thanks{Online video material available at \texttt{http://www.tut.fi/5G/TWC16/}.}

\markboth{ }%
{ }

\maketitle

\vspace{-25pt}

% Smaller space after the authors
%\vspace{-40pt}

{\linespread{\speciallinespacing}
\let\thefootnote\relax\footnotetext{
\footnotesize
%M.~Koivisto, A. Hakkarainen, and M.~Valkama are with the Department of Electronics and Communications Engineering, Tampere University of Technology, FI-33101 Tampere, Finland (email: mike.koivisto@tut.fi).

This work was supported by the Finnish Funding Agency for Technology and Innovation (Tekes), under the projects \textquotedblleft{5G Networks and Device Positioning\textquotedblright}, and \textquotedblleft{Future Small-Cell Networks using Reconfigurable Antennas\textquotedblright}.

On-line videos available at \texttt{\url{http://www.tut.fi/5G/GLOBECOM16}}

This work has been submitted to the IEEE for possible publication. Copyright may be transferred without notice, after which this version may no longer be accessible.

}}

\begin{abstract}
It is commonly expected that future \gls{5G} networks will be deployed with a high spatial density of \glspl{AN} in order to meet the envisioned capacity requirements of the upcoming wireless networks. Densification is beneficial not only for communications but it also creates a convenient infrastructure for highly accurate \gls{UN} positioning. Despite the fact that positioning will play an important role in future networks, thus enabling a huge amount of location-based applications and services, this great opportunity has not been widely explored in the existing literature. Therefore, this paper proposes an \gls{UKF}-based method for estimating \DOAs and \TOAs at \glspl{AN} as well as performing joint 3D positioning and network synchronization in a network-centric manner. In addition to the proposed \gls{UKF}-based solution, the existing 2D \gls{EKF}-based solution is extended to cover also realistic 3D positioning scenarios. Building on the premises of \gls{5G} \glspl{UDN}, the performance of both methods is evaluated and analysed in terms of \DOA and \TOA estimation as well as positioning and clock offset estimation accuracy, using the METIS map-based ray-tracing channel model and 3D trajectories for vehicles and \glspl{UAV} through the Madrid grid. Based on the comprehensive numerical evaluations, both proposed methods can provide the envisioned one meter 3D positioning accuracy even in the case of unsynchronized \gls{5G} network while simultaneously tracking the clock offsets of network elements with a nanosecond-scale accuracy.
\end{abstract}

\glsresetall

\vspace{-5pt}
\vspace{-10pt}

% \fi
\IEEEpeerreviewmaketitle

%!TEX root = main.tex

\section{Introduction} \label{sec:intro}

In contrast to earlier generations, the structure of future \gls{5G} networks will change dramatically in order to meet the envisioned requirements in terms of, e.g., high data rates and latency. To achieve these demanding requirements, it is commonly expected that \gls{5G} networks will be deployed with wideband waveforms at high, even above \SI{6}{GHz} frequencies due to better spectrum availability\cite{osseiran_scenarios_2014, boccardi_five_2014,5g-ppp_2015}. Furthermore, it is envisioned that spatial density of \glspl{AN} will simultaneously increase resulting in so-called \glspl{UDN} \cite{ngmn_alliance_5g_2014, 5g-ppp_2015, 5g_forum_5g_2015}. Thus, \glspl{UN} will be most likely within the range of several \glspl{AN} leading to a scenario where devices will also be in a \gls{LoS} condition with a few \glspl{AN} most of the time\cite{dammann_prospects_2015, metis_channel_2015}. Such a condition is beneficial not only for communication purposes but it also creates a great opportunity for device positioning based on highly accurate \gls{ToA} estimates that can be obtained at \gls{LoS}-\glspl{AN} due to the envisioned wideband waveforms. In addition, \glspl{AN} in \gls{5G} are expected to be equipped with smart antenna solutions, such as antenna arrays, enabling also estimation of \glspl{DoA}. This directional information can be, in turn, fused with \gls{ToA} estimates to achieve even improved positioning performance. 

Technically, positioning in \gls{5G} networks can be carried out either within devices or in a network-centric manner, where the central unit of a network tracks the devices within the network. However, the latter option has several advantages over a device-centric approach. First, network-centric positioning is more energy efficient from the devices' perspective since the actual computational effort is done in the central units of networks, thus saving the batteries of \glspl{UN}. Furthermore, \gls{UL} pilot signals, which are anyway exchanged between the \glspl{UN} and \glspl{AN} for channel estimation and scheduling on \gls{TDD} networks~\cite{kela_borderless_2015}, can be utilized for highly accurate positioning without the need of allocating specific reference signals for positioning~\cite{werner_joint_2015}. This means, in turn, that such positioning engine can be continuously running in the background providing up-to-date location information for necessary systems and applications with a low latency. 

In general, \gls{5G} has a great potential to achieve the envisioned positioning accuracy of one meter or even less~\cite{5g-ppp_2015}, consequently outperforming existing positioning techniques such as \gls{GPS} and \gls{WLAN} fingerprinting based methods, in which the positioning accuracy is usually in the order of couple of meters~\cite{dardari_indoor_2015}. Such a significant improvement in positioning accuracy creates an opportunity for huge amount of future location based applications, e.g., \glspl{ITS}, autonomous vehicles, and proactive \gls{RRM}~\cite{di_taranto2014}. Despite the fact that positioning will naturally play an important role in future \gls{5G} networks, the potential for positioning in such networks has not been widely recognized topic in the existing literature.

%In order to estimate and track positions of devices, an iterative estimation method need to be applied to the known system models. 
%Due to the nature of assumed and potential \gls{DoA} and \gls{ToA} measurements, the well-known \gls{KF} is not suitable filtering method for the desired positioning system. However, one widely used method for these estimation problems is the \gls{EKF} in which non-linear models are approximated using first-order Taylor approximations before applying the \gls{KF} equations. 
In the past, the \glspl{EKF} have been proposed for 2D target tracking using \gls{DoA} measurements only~\cite{aidala_kalman_1979}, but also a combination of \gls{DoA} and \gls{ToA} measurements is used for tracking devices in a synchronized network~\cite{navarro_toa_2007}. In addition, the authors in~\cite{werner_joint_2015} proposed a method for joint device positioning and synchronization using \glspl{DoA} and \glspl{ToA} in future \gls{5G} networks in order to relax a synchronization assumption. Furthermore, realism was increased in~\cite{koivisto_joint_2016} by assuming not only an unsynchronized \gls{UN} but also unsynchronized \glspl{AN} in an \gls{EKF}-based joint 2D positioning and network synchronization method. In contrast to the proposed 2D positioning solutions, in~\cite{irahhauten_joint_2012} a joint \gls{DoA} and \gls{ToA} based method also for 3D device location estimation using only one \gls{AN} in a synchronized network was proposed. However, the authors in~\cite{irahhauten_joint_2012} did not propose any iterative filtering method to obtain continuous position estimates for the devices.

In this paper, an \gls{UKF}-based cascaded solution for the joint \gls{DoA} and \gls{ToA} estimation, and simultaneous 3D device positioning and network synchronization in future \gls{5G} \glspl{UDN} is proposed. In the first phase of the proposed cascaded solution, a \gls{UKF}-based estimation and tracking method for both \glspl{DoA} and \glspl{ToA}, which stems from~\cite{koivisto_joint_2016}, is performed. Thereafter, the obtained estimates are fused within the joint 3D device positioning and network synchronization \gls{UKF}, later referred to as the joint \POSCLOCKUKF or joint \POSSYNCUKF depending on whether the \glspl{AN} composing the network are assumed to be synchronized among each other or phase-locked, respectively. As a result, the proposed method provides not only 3D position estimates for a \gls{UN} but also clock offset estimates for the \gls{UN} and \gls{LoS}-\glspl{AN}. In addition to the \gls{UKF}-based approach, the existing 2D \gls{EKF}-based positioning method proposed in~\cite{koivisto_joint_2016} is extended to 3D positioning and network synchronization. In contrast to earlier studies, the proposed methods provide highly accurate and sequential 3D position estimates for the \glspl{UN}, thus enabling tracking of \glspl{UAV}, for example.

The rest of the paper is organized as follows. In Section \ref{sec:system}, necessary assumptions about the network, used positioning engine as well as channel model are discussed. Thereafter, a general algorithm for the \gls{UKF} is presented in Section \ref{sec:ukf}, and extended for \gls{DoA} and \gls{ToA} estimation, as well as simultaneous \gls{UN} positioning and network synchronization in Sections \ref{sec:doatoaukf} and \ref{sec:posukf}, respectively. In Section \ref{sec:results}, the employed simulation environment is explained together with the assumptions made before the actual numerical results and analysis of the proposed method are presented. Finally, conclusions are drawn in Section~\ref{sec:conclusion} in order to summarize the main contributions of this work.

%\begin{itemize}
%\item 3D instead of 2D (enables, e.g., UAV tracking)
%\item synchronization as a by-product (ANs are also unsynchronized meaning that there exists a clock offset in the AN clo clock)
%\item Vehicle attached device
%\item motion model in \cite{werner_joint_2015} is not probably the best possible (comparison in the results?)
%\end{itemize}

%!TEX root = main.tex

\section{System Model}\label{sec:system}

\subsection{\gls{5G} Network and Positioning Engine}

\begin{figure}[!t]
    \centering
    \includegraphics[width=3.35in]{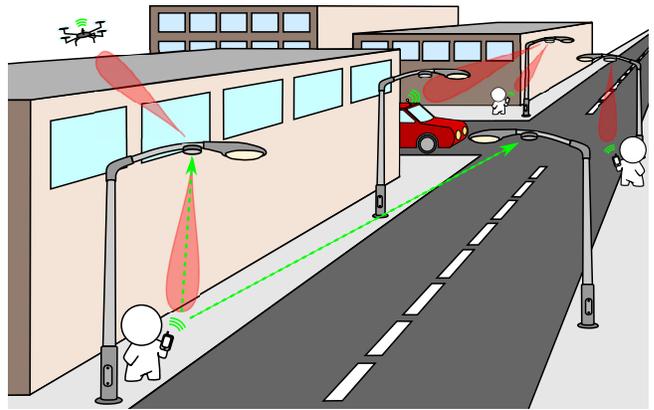}
    \caption{Densely deployed \glspl{AN} attached to lamp posts in an outdoor \gls{UDN} with several connected devices. In the proposed method, \gls{UN} positioning and network synchronization are performed in a network-centric manner in order to reduce energy consumption of the devices.}
    \label{fig:udn}
    \vspace{\spaceunderfig}
\end{figure}

In this paper, following the key \gls{5G} white papers\cite{5g_forum_5g_2015,ngmn_alliance_5g_2014, 5g-ppp_2015}, an outdoor network with densely deployed \glspl{AN} is considered as illustrated in Fig.~\ref{fig:udn}. In the assumed network, the \glspl{AN} are attached to lamp posts \SI{7}{m} above the ground and with an \gls{ISD} of around \SI{50}{m} or even less. The deployed \glspl{AN} are also assumed to be equipped with antenna arrays which enable \gls{DoA} estimation at \glspl{AN}. In the considered network, the \gls{AN} consist of cylindrical antenna arrays in which 10 dual-polarized cross-dipoles are placed along two circles. However, other multiantenna solutions can be used as well. For the sake of simplicity, the locations of \glspl{AN}, denoted as $\vec{p}_{\ell_i} = (x_{\ell_i},y_{\ell_i},z_{\ell_i})$ where $\ell_i$ denotes the index of an \gls{AN}, are assumed to be known using, e.g., information from \gls{GPS}. 

%\frame{\includegraphics[width=3.36in]{figures/udn5}}

\begin{figure}[!t]
    \centering
    \vspace{5pt}
    \def\svgwidth{3.5in}
    \textrm{\footnotesize{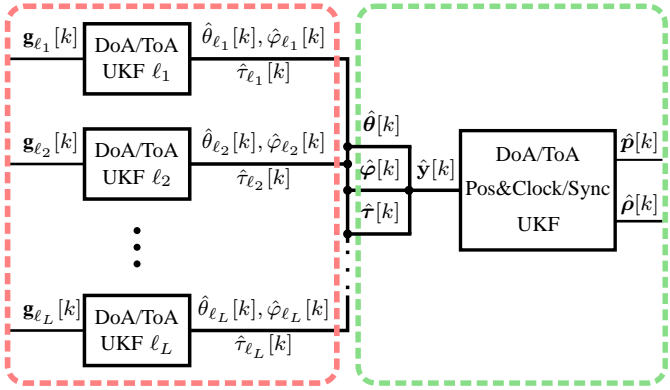}}
    \caption{In the \gls{DoA} and \gls{ToA} estimation phase (red color), channel estimates $\vec{g}$ are used within the \DOATOAUKFs at each \LOSAN to obtain azimuth and elevation angle as well as \gls{ToA} estimates denoted as $\hat{\varphi}$ and $\hat{\theta}$, and $\hat{\tau}$, respectively. The estimates are then fused within the joint \gls{DoA}/\gls{ToA} Pos\&Clock/Sync \gls{UKF} into a \gls{UN} position estimate $\hat{\vec{p}}$ and clock offset estimates $\hat{\boldsymbol{\rho}}$ in the second phase of the proposed solution (green color).
    %In the first phase of the cascaded \gls{UKF} solution (rounded with red), azimuth and elevation angles denoted as $\varphi$ and $\theta$, respectively, as well as \glspl{ToA} $\tau$ are estimated and tracked at \glspl{AN} using \gls{DoA}/\gls{ToA} \glspl{UKF}. The obtained estimates are fused into measurements $\hat{\vec{y}}$ in the joint 3D \gls{DoA}/\gls{ToA} Pos\&Clock/Sync \gls{UKF} method in the second phase of the proposed solution (rounded with green).}
    }
    \label{fig:cascaded_ukf}
    \vspace{\spaceunderfig}
\end{figure}

In the considered network, \glspl{UN} periodically transmit \gls{UL} pilot signals which will be later referred to as beacons. The beacons are assumed to employ an \gls{OFDM} waveform, in the form of \gls{OFDMA} in a multiuser network. The beacons are used to obtain \gls{CSI} at \glspl{AN}~\cite{kela_borderless_2015} but these beacons can be also utilized for network-centric positioning, thus  leading to an ''always-on'' positioning solution. 
%Therefore, dedicated reference signals for positioning purposes may not be required in future \gls{5G}. 
Based on the received beacons, the \glspl{AN} will detect whether or not they are in \gls{LoS} condition with a \gls{UN}. Such a \gls{LoS} condition can be determined using, e.g., Rice-K factor of the \gls{RSS}~\cite{BGTV07}. Each \gls{LoS}-\gls{AN} will then estimate \glspl{DoA} and \glspl{ToA} using the received beacons within the \DOATOAUKFs, and these estimates are thereafter gathered from all \gls{LoS}-\glspl{AN} and fused into a 3D \gls{UN} position estimate in the central entity of the network using the proposed joint \gls{DoA}/\gls{ToA} Pos\&Clock or Pos\&Sync \gls{UKF} as depicted in Fig.~\ref{fig:cascaded_ukf}.

\subsection{Clock Models}

In general, the clocks within devices are equipped with relatively cheap oscillators which, in turn, leads to time-varying clock offsets in such devices. Due to these imperfections, a progressive model is required in order to characterize or estimate clock offsets of the devices. It is shown in\cite{kim_tracking_2012} that clock offset $\rho$ for two consecutive time instants $k-1$ and $k$ are related such that
\begin{align}
    \rho [k] = \rho [k-1] + \alpha [k] \Delta t,
\end{align}
where $\Delta t$ is the time period between the instants $k-1$ and $k$, and $\alpha[k]$ denotes the clock skew which is a time-varying quantity as well. Based on observations, e.g., in~\cite{kim_tracking_2012}, the average clock skew can be considered constant but in this paper, the following time-dependent clock skew model that stems from~\cite{kim_tracking_2012} and encompasses also the constant clock skew model as a special case is considered
\begin{align}
    \alpha[k] = \beta \alpha[k-1] + \eta [k],
\end{align}
where $\eta[k] \sim \mathcal{N}(0,\sigma_{\eta}^2)$ is an additive Gaussian white-noise sequence and $\beta$ is a constant model parameter such that $| \beta | < 1$.

Two different scenarios for synchronization within a network are considered. In the first scenario, \glspl{UN} have unsynchronized clocks where only the necessary timing and frequency synchronization to avoid \gls{ICI} and \gls{ISI} is assumed. On the other hand, \glspl{AN} in such a scenario are assumed to have synchronized clocks among each other. In the second and more realistic scenario, clocks within \glspl{AN} are set to phase-locked, i.e., the clock offsets of \glspl{AN} are not fundamentally varying over the real time, whereas clocks within \glspl{UN} are assumed to be unsynchronized. The aforementioned synchronized and phase-locked clocks within \glspl{AN} can be obtained using a time reference from, e.g., GPS, or by communicating a reference time from the central unit of the network to the \glspl{AN}. However, these methods surely increase the signaling overhead.
 
%In this paper, two different scenarios for synchronization within a network are assumed. In the first scenario, \glspl{UN} are assumed to have unsynchronized clocks when necessary timing and frequency synchronization to avoid \gls{ICI} and \gls{ISI} is achieved. On the other hand, \glspl{AN} in such scenario are assumed to have synchronized clocks among each other. In the second and more realistic scenario, clocks within \glspl{AN} are set to phase-locked, i.e., the clock offsets of \glspl{AN} are not fundamentally varying over the real time, whereas clocks within \glspl{UN} are assumed to be unsynchronized. Aforementioned synchronized and phase-locked clocks within \glspl{AN} can be obtained using a time reference from, e.g., GPS, or by communicating a reference time from a central unit of the network to the \glspl{AN}. However, these methods surely increase the signaling overhead. 

\subsection{Channel Model}

In order to perform \DOA and \TOA estimation at \LOSANs  as presented later in Section~\ref{sec:doatoaukf}, the following \gls{SIMO} multiantenna-multicarrier channel response model for $\mathcal{M}_{\textrm{AN}}$ antenna elements and $\mathcal{M}_{f}$ subcarriers is exploited~\cite{Ric05}
\begin{align}
    \vec{g}_{\ell_k} \approx \mat{B}_{\ell_k}(\theta,\varphi, \tau)\vec{\gamma} + \vec{n},
    \label{eq:channel_model}
\end{align}
where $\vec{g}_{\ell_k} \in \mathbb{C}^{\mathcal{M}_{\textrm{AN}}\mathcal{M}_{f}}$ is a channel response vector obtained at the \LOSAN with an index $\ell_k$. Furthermore, $\mat{B}_{\ell_k}(\theta,\varphi,\tau) \in \mathbb{C}^{\mathcal{M}_{\textrm{AN}}\mathcal{M}_{f} \times 2}$ and $\vec{\gamma} \in {\mathbb{C}^{2}}$ denote the polarimetric response of the multiantenna \gls{AN} $\ell_k$ and complex path weights, respectively. The model \eqref{eq:channel_model} is perturbed with complex-circular zero-mean white-Gaussian noise $\vec{n} \in \mathbb{C}^{\mathcal{M}_{\textrm{AN}}\mathcal{M}_{f}}$ with variance $\sigma_n^2$. 

The polarimetric antenna array response is given in terms of the \gls{EADF} such that 
\begin{equation}
    \begin{aligned} 
    \mathbf{B}_{\ell_k}(\theta,\varphi,\tau) = [&\mathbf{G}_H \mathbf{d}(\varphi,\theta) \otimes \mathbf{G}_f \mathbf{d}(\tau),\\
    &\mathbf{G}_V \mathbf{d}(\varphi,\theta) \otimes \mathbf{G}_f \mathbf{d}(\tau)],
    \label{eq:multichannel_model}
    \end{aligned}
\end{equation}
where $\otimes$ denotes the Kronecker product, and  $\mathbf{G}_H \in \mathbb{C}^{\mathcal{M}_{\textrm{AN}} \times \mathcal{M}_{a} \mathcal{M}_{e}}$ and $\mathbf{G}_V \in \mathbb{C}^{\mathcal{M}_{\textrm{AN}} \times \mathcal{M}_{a} \mathcal{M}_{e}}$ are the \gls{EADF} for horizontal and vertical excitations, respectively. Numbers of the determined array response modes, i.e., spatial harmonics, in \gls{EADF}~\cite{Ric05} are denoted as $\mathcal{M}_{a}$ and $\mathcal{M}_{e}$ for azimuth and elevation, respectively. Furthermore, $\mathbf{G}_f \in \mathbb{C}^{\mathcal{M}_f \times \mathcal{M}_f}$ is the frequency response of the transceiver, and $\mathbf{d}(\varphi,\theta) \in \mathbb{C}^{\mathcal{M}_{a}\mathcal{M}_{e}}$ can be written as 
\begin{align}
  \mathbf{d}(\varphi,\theta) = \mathbf{d}(\theta) \otimes \mathbf{d}(\varphi)    
\end{align}
%$\mathbf{d}(\theta,\varphi) = \mathbf{d}(\varphi) \otimes \mathbf{d}(\theta) \in \mathbb{C}^{\mathcal{M}_{a} \mathcal{M}_{e} \times 1}$, 
where $\mathbf{d}(\varphi) \in \mathbb{C}^{\mathcal{M}_a}$ and $\mathbf{d}(\theta) \in \mathbb{C}^{\mathcal{M}_e}$ as well as $\vec{d}(\tau)\in \mathbb{C}^{\mathcal{M}_f}$ in \eqref{eq:multichannel_model} are Vandermonde structured vectors. These vectors map the spatial and temporal parameters to the relative frequency domain such that
\begin{align}
    \mathbf{d}(\tau) = \left[ e^{-j\pi(\mathcal{M}_f -1)f_0 \tau}, \dots, 1, \dots,  e^{j\pi(\mathcal{M}_f -1) f_0\tau} \right] \transpose,
\end{align}
which can be transformed to correspond the similar $\vec{d}(\varphi)$ and $\vec{d}(\theta)$ using the relation $\varphi/2 = \pi f_0 \tau$, where $f_0$ denotes the subcarrier spacing of the employed \gls{OFDM} waveform. The array calibration data, represented using the \gls{EADF}, can be determined in well-defined propagation environments, e.g., in anechoic chamber~\cite{Ric05}. In this paper, the \glspl{EADF} are assumed to be known for all \glspl{AN}.

%\begin{itemize}
%\item Physical: Network, ISD, Antenna model, UL signals
%\item Cascaded solution using UKFs at ANs and in the fusion
%\item abbreviations for the methods
%\end{itemize}

%!TEX root = main.tex

\section{Bayesian Filtering Methods} \label{sec:ukf}

%A noisy state of a system where both state evolution and measurement model are linear can be estimated recursively using the well-known \gls{KF}. However, these models that describes the system of interest are usually non-linear instead and hence it is not possible to apply the \gls{KF} equations to the estimation problems as such. 
In this paper, two filtering methods for sequential state estimation of non-linear systems, namely \gls{EKF} and \gls{UKF} are used. Let us first denote the state of a system at time step $k$ as $\vec{s}[k] \in \mathbb{R}^n$. Moreover, an additive and linear state transition model that describes the state evolution between two consecutive time instants $k-1$ and $k$ is assumed such that
\begin{align}
\vec{s}[k] &= \mat{F}\vec{s}[k-1] + \vec{u}[k],
\label{eq:general_state_model}
\end{align}
where $\mat{F}\in \mathbb{R}^{n \times n}$ denotes a state transition matrix, and $\vec{u}[k] \sim \mathcal{N}(0,\mat{Q}_k)$ is a zero-mean Gaussian process noise with covariance $\mat{Q}_k \in \mathbb{R}^{n \times n}$. Furthermore, a measurement model that relates the current state $\vec{s}[k]$ to the obtained measurements $\vec{y}[k] \in \mathbb{R}^m$ through a non-linear function $\mathbf{h}: \mathbb{R}^n \rightarrow \mathbb{R}^m$ is expressed as  
\begin{align}
\vec{y}[k] &= \mathbf{h}(\vec{s}[k]) + \vec{w}[k],
\label{eq:general_meas_model}
\end{align}
where $\vec{w}[k] \sim \mathcal{N}(0,\mat{R}_k)$ is a zero-mean Gaussian measurement noise with covariance $\mat{R}_k\in \mathbb{R}^{m \times m}$. In addition to the models \eqref{eq:general_state_model} and \eqref{eq:general_meas_model}, also the initial prior distribution of the state $\vec{s}[0] \sim \mathcal{N}(\vec{m}_{0}, \mat{P}_{0})$ is assumed to be known. In the following section, a general algorithm for the \gls{UKF} is shortly presented while the algorithm for the well-known \gls{EKF} can be found, e.g., in~\cite{sarkka2013}.

\subsection{Unscented Kalman Filter} 

%A noisy state of a system where both state evolution and measurement model are linear can be estimated recursively using the well-known \gls{KF}. However, the models that describes the system of interest are usually non-linear instead and hence it is not possible to apply the \gls{KF} to such estimation problems. 
%One possible and widely used solution to overcome these issues is the \gls{EKF} in which non-linear models are linearized using first-order Taylor approximations so that the general \gls{KF} equations can be applied\REFC.
%Despite the effectiveness of the \gls{EKF}, it has couple of drawbacks. First, linear approximations within the \gls{EKF} can not fully capture the behaviour of a non-linear system even resulting in an unstable filter. Second, the derivation of the necessary Jacobian matrices are often non-trivial and prone to errors leading to significant implementation difficulties\REFC. 
%In the following part of the paper, an algorithm of the \gls{UKF} is shortly presented. 
The \gls{UKF} is an estimation algorithm which uses deterministic sample points, i.e., sigma points to address the limitations of the \gls{EKF} to some extent. Instead of approximating the non-linear models, the generated sigma points are propagated through the involved non-linearities in order to approximate the final posterior distribution~\cite{julier_new_1995}. Using this so-called unscented transformation based approximation, high-order information about the desired distribution can be expressed with a relatively small number of fixed points~\cite{julier_unscented_2004}. A compact representation of the \gls{UKF} in its general form is shown in Algorithm~\ref{alg:alg_ukf}, and it follows mainly the same notation as in~\cite{sarkka2013}.

The algorithm of the \gls{UKF} consists of prediction and update phases, in which the non-linearities of the state and measurement models are captured by propagating the deterministic sigma points according to the models. In addition to known initial distribution $\vec{s}[0] \sim \mathcal{N}(\vec{m}_{0}, \mat{P}_{0})$, a scaling parameter $\lambda$, which can be expressed in terms of algorithm parameters $\alpha$ and $\kappa$, needs to be defined as
\begin{align}
\lambda = \alpha^2(n+\kappa)-n,
\label{eq:lambda}
\end{align}
where $n$ denotes the dimension of the state, and the parameters $\alpha$ and $\kappa$ are used to determine the spread of the sigma points around the mean. In literature, the parameter $\alpha$ is usually set to a small positive value, e.g., $1 \leq \alpha \leq 1e^{-5}$, and values of $0$ and $3-n$ are commonly used for the secondary algorithm parameter $\kappa$~\cite{wan_unscented_2000, sarkka2013, julier_new_1995, julier_unscented_2004}.

Generated sigma points are associated with the corresponding weights which can be evaluated such that
\begin{equation}
\begin{aligned}
&W^{(0)}_m = \frac{\lambda}{n+\lambda}, && W^{(i)}_m = \frac{1}{2(n+\lambda)}, \\
&W^{(0)}_c = \frac{\lambda}{n+\lambda} + (1-\alpha^2 + \beta), && W^{(i)}_c = \frac{1}{2(n+\lambda)},
\end{aligned}
\label{eq:weights}
\end{equation}
where $\beta$ is an additional algorithm parameter that can be used to incorporate prior information of the state. For the Gaussian priors, the optimal choice is $\beta=2$~\cite{wan_unscented_2000}. 

\begin{algorithm}
\caption{Unscented Kalman Filter}
\label{alg:alg_ukf}
\textbf{Initialization} Define initial distribution $\vec{s}[0] \sim \mathcal{N}(\vec{m}_{0}, \mat{P}_{0})$\\

\textbf{For} every iteration $k = 1, \dots, T$
\begin{enumerate}
\item Due to the considered linear state model~\eqref{eq:general_state_model}, the \textit{a priori} estimates of the mean and covariance can be calculated as
\begin{align}
\vec{m}_{k}^- &= \mat{F}\vec{m}_{k-1},\\
\mat{P}_{k}^- &= \mat{F}\mat{P}_{k-1}\mat{F}\transpose + \mat{Q}_{k}.
\end{align}
%\item Generate $2n + 1$ sigma points according to 
%\begin{equation}
%\begin{aligned}
%\mathcal{X}_{k-1}^{(0)} &= \vec{m}_{k-1}\\
%\mathcal{X}_{k-1}^{(i)} &= \vec{m}_{k-1} + \sqrt{\lambda + n} \left[ \sqrt{\mat{P}_{k-1}}\right]_i \\%&& i=1,\dots,n\\
%\mathcal{X}_{k-1}^{(n+i)} &= \vec{m}_{k-1} - \sqrt{\lambda + n} \left[ \sqrt{\mat{P}_{k-1}}\right]_{i}, %&& i=1,\dots,n,
%\end{aligned}
%\label{eq:sigmas_1}
%\end{equation}
%where $i = 1,\dots, n$.

%\item Propagate the sigma points according to the state model%~\eqref{eq:general_state_model}
%\begin{align}
%\hat{\mathcal{X}}_{k}^{(i)} = \mat{F}\mathcal{X}_{k-1}^{(i)}, && i = 0,1,\dots,2n.
%\end{align}

%\item Compute the predicted mean and covariance
%\begin{align}
%\mathbf{m}^-_k &= \sum_{i=0}^{2n} W^{(i)}_m \hat{\mathcal{X}}^{(i)}_{k} \\
%\mathbf{P}^-_k &= \sum_{i=0}^{2n} W^{(i)}_c (\hat{\mathcal{X}}^{(i)}_{k}-\mathbf{m}^-_k)(\hat{\mathcal{X}}^{(i)}_{k}-\mathbf{m}^-_k)\transpose + \mat{Q}_k.
%\end{align}

\item Form $2n+1$ sigma points according to
\begin{equation}
\begin{aligned}
\mathcal{X}_{k}^{-(0)} &= \vec{m}^-_{k}\\
\mathcal{X}_{k}^{-(i)} &= \vec{m}^-_{k} + \sqrt{\lambda + n} \left[ \sqrt{\mat{P}^-_{k}}\right]_i\\% && i=1,\dots,n\\
\mathcal{X}_{k}^{-(n+i)} &= \vec{m}^-_{k} - \sqrt{\lambda + n} \left[ \sqrt{\mat{P}^-_{k}}\right]_{i}, %&& i=1,\dots,n,
\end{aligned}
\label{eq:sigmas_2}
\end{equation}
where $i = 1,\dots, n$.

\item Propagate the sigma points through the measurement model~\eqref{eq:general_meas_model}
\begin{align}
&\hat{\mathcal{Y}}^{(i)}_k = \mathbf{h}(\mathcal{X}^{-(i)}_k), && i = 0,1,\dots,2n,
\end{align}

\item Compute the measurement mean, measurement covariance and cross-covariance
\begin{align}
\vec{\mu}_k &= \sum_{i=0}^{2n} W^{(i)}_m \hat{\mathcal{Y}}^{(i)}_k, \\
\mat{S}_k &= \sum_{i=0}^{2n} W^{(i)}_c (\hat{\mathcal{Y}}^{(i)}_k-\vec{\mu}_k)(\hat{\mathcal{Y}}^{(i)}_k-\vec{\mu}_k)\transpose + \mat{R}_k, \\
\mat{C}_k &= \sum_{i=0}^{2n} W^{(i)}_c (\mathcal{X}^{-(i)}_k-\vec{m^-})(\hat{\mathcal{Y}}^{(i)}_k-\vec{\mu}_k)\transpose.
\end{align}

\item Compute the \textit{a posteriori} mean and covariance of $\vec{s}[k] \sim \mathcal{N}(\vec{m}_{k}, \mat{P}_{k})$  using the measurements $\vec{y}[k]$
\begin{align}
\mathbf{K}_k &= \mat{C}_k\mat{S}^{-1}_k\\
\vec{m}_k &= \vec{m}^-_k + \mat{K}_k(\vec{y}[k]-\vec{\mu}_k)\\
\mat{P}_k &= \mat{P}^-_k - \mat{K}_k\mat{S}_k\mat{K}_k\transpose,
\end{align}

\end{enumerate}
\textbf{End}
\end{algorithm}
%!TEX root = main.tex

\section{UKF-based DoA and ToA Estimation at ANs} \label{sec:doatoaukf}

In this section, the first phase of the proposed cascaded \gls{UKF} based solution depicted in Fig.~\ref{fig:cascaded_ukf} is shortly described. The proposed method follows the \gls{EKF}-based method proposed earlier in~\cite{koivisto_joint_2016}. In these \DOATOAUKFs, only a single propagation path with the highest power, which usually corresponds to the \gls{LoS} path, is tracked. Let us assume the following state for the \DOATOAUKFs 
\begin{align}
\vec{s}[k] = \left[ \tau[k],\, \varphi[k],\, \theta[k],\, \Delta\tau[k],\, \Delta\varphi[k], \, \Delta\theta[k] \right],
\end{align}
where $\tau,\, \varphi$, and $\theta$ denote \gls{ToA}, azimuth \gls{DoA} and elevation \gls{DoA}, respectively. Furthermore, parameters $\Delta\tau,\, \Delta\varphi$, and $\Delta\theta$ correspond to the rate-of-change of \gls{ToA} and \gls{DoA} angles. Considering a linear \gls{CV} model, the state transition matrix $\mat{F} \in \mathbb{R}^{6 \times 6}$ and the covariance $\mat{Q}_k \in \mathbb{R}^{6 \times 6}$ of the state noise $\vec{u}[k] \sim \mathcal{N}(0,\mat{Q}_k)$ in~\eqref{eq:general_state_model} can be written such that\renewcommand{\arraystretch}{1.4}
\begin{align}
    \mat{F} = \left[ \begin{array}{cc} 
                     \mat{I}_{3\times3} & \Delta t  \mat{I}_{3 \times 3}\\ \mat{0}_{3 \times 3} & \mat{I}_{3\times3}\end{array} \right], && \mat{Q}_k = \begin{bmatrix}
            \frac{\Delta t^3 \mat{D}}{3}   & \frac{\Delta t^2  \mat{D}}{2}\\
            \frac{\Delta t^2 \mat{D}}{2}& \Delta t \mat{D}
        \end{bmatrix},
\end{align}
respectively.
%the state noise $\vec{w}[k] \sim \mathcal{N}(0,\mat{Q}_k)$ and $\Delta t$ denotes time difference between the consecutive time steps $k-1$ and $k$. Furthermore, the state noise $\vec{w}[k] \sim \mathcal{N}(0,\mat{Q}_k)$, where 
%\begin{align}
%    \mat{Q}_k = \begin{bmatrix}
%            \frac{\Delta t^3 \cdot \mat{D}}{3}   & \frac{\Delta t^2\cdot \mat{D}}{2}\\
%            \frac{\Delta t^2\cdot \mat{D}}{2}& \Delta t \cdot \mat{D}
%        \end{bmatrix} \in \mathbb{R}^{6 \times 6},
%\end{align}
Here, $\mat{D} = \diag{(\sigma_{\tau}^2,\sigma_{\varphi}^2,\sigma_{\theta}^2)} \in \mathbb{R}^{3 \times 3}$ is a diagonal matrix consisting of variances of the rate-of-change state parameters, and $\Delta t$ denotes the time difference between the consecutive time steps $k-1$ and $k$.

The actual state parameter estimation is performed in the \DOATOAUKFs mainly according to the \gls{UKF} algorithm in Algorithm~\ref{alg:alg_ukf}. Since the state parameters $\varphi$, $\theta$, and $\tau$ are defined on a circle, the generated sigma points are mapped to the feasible regions after carrying out \eqref{eq:sigmas_2} such that $\varphi \in (0,2\pi]$ and $\theta \in [0,\pi]$. Furthermore, due to the complexity of the estimation problem and complex-valued data, the measurement update phase of the \gls{UKF}, i.e., steps 6 and 7 in Algorithm~\ref{alg:alg_ukf}, is performed using a sequential Gauss-Newton method leading to the \textit{a posteriori} mean and covariance estimates such that 
\begin{align}
    \vec{m}_k &= \vec{m}^-_k + \mat{J}^{-1}\vec{r}(\vec{m}^-_k),\\
    \mat{P}_k &= \mat{J}^{-1}.
\end{align}
Here, $\mat{J}^{-1}$ denotes an approximation of the inverse of the empirical \gls{FIM} obtained using weighted statistical linear regression, and $\vec{r}$ denotes a cost function gradient evaluated at the \textit{a priori} mean $\vec{m}^-_k$. More details regarding such an alternative form of the \gls{UKF} can be found in~\cite{Merwe04}. Finally, the \DOATOAUKFs at \glspl{AN} can be initialized according to the efficient approach proposed for the similar \gls{EKF} based  method in~\cite{koivisto_joint_2016}
%!TEX root = main.tex

\section{Proposed UKF-Based Joint 3D Positioning and Synchronization} \label{sec:posukf}

In the second phase of the proposed cascaded solution, the estimated \glspl{DoA} and \glspl{ToA} that are communicated to the central entity of a network are, thereafter, fused within the joint positioning and synchronization method as depicted in Fig.~\ref{fig:cascaded_ukf}. In this section, the models used for the 3D positioning and synchronization purposes for both \gls{UKF} and generalized \gls{EKF} based \DOA/\TOA Pos\&Clock and \DOA/\TOA Pos\&Sync filters are presented.

%, \,\vec{\rho}_{\ell_{\textrm{AN}}}
The state of the 3D \POSCLOCKUKF and \gls{EKF} for the considered \gls{CV} motion model in the case of synchronized \glspl{AN} can be written as 
\begin{align}
\vec{s}[k] &= \left[ \vec{p}\transpose[k],\,\vec{v}\transpose[k],\, \rho_{\textrm{UN}}[k],\, \alpha[k]\right]\transpose \in \mathbb{R}^{8},
%\vec{s}[k] &= \left[ \vec{p}[k],\,\vec{v}[k],\,\vec{a}[k],\, \rho_{\textrm{UN}}[k],\, \alpha[k]\right]\transpose,
%\vec{s}[k] &= \left[ \vec{p}[k],\,\vec{v}[k],\, \omega[k],\, \rho_{\textrm{UN}}[k],\, \alpha[k]\right]\transpose,
\end{align}
where $\vec{p}[k] = [x[k],y[k],z[k]]\transpose$ is the 3D position and $\vec{v}[k] = [v_x[k], v_y[k],v_z[k]]\transpose$ 
%$\vec{a}[k] = [a_x[k], a_y[k],a_z[k]]\transpose$ 
is the 3D velocity
%, and acceleration 
of a \gls{UN}. Furthermore, $\rho_{\textrm{UN}}[k]$ and $\alpha[k]$ denote the clock offset and clock skew of the \gls{UN} at time step $k$, respectively. In the scenario where the \glspl{AN} are  phase-locked, the mutual clock offsets of \LOSANs need to be augmented to the state of the \POSSYNCUKF and \gls{EKF} such that $
%\begin{align}
\vec{s}[k] \leftarrow \left[ \vec{s}\transpose[k],\,\vec{\rho}_{\ell_{\textrm{AN}}}\transpose[k]\right]\transpose,$
%\end{align}
where $\vec{\rho}_{\ell_{\textrm{AN}}}[k] = [\rho_{\ell_1}[k], \dots, \rho_{\ell_{L}}[k]]\transpose$ consists of the clock offsets of all $L$ \glspl{AN} which are in \gls{LoS} condition with the \gls{UN} at time step $k$. Note that all the clock offsets are with respect to a chosen reference \gls{AN}.

%In order to perform positioning and clock offset estimation of a UN and \gls{LoS}-\glspl{AN} jointly within the joint \POSSYNCUKF, the following state is defined 
%\begin{align}
%\vec{s}[k] = \left[ \vec{p}[k],\,\vec{v}[k],\, \omega[k],\, \rho_{\textrm{UN}}[k],\, \alpha[k], \,\vec{\rho}_{\ell_{\textrm{AN}}}\right]\transpose,
%\end{align} 
%where $\vec{p}[k] = [x[k],\, y[k],\, z[k]]\transpose$, $\vec{v}[k] = [v_x[k],\, v_y[k],\, v_z[k]]\transpose$, and $\omega[k]$ are three-dimensional position, velocity, and turn rate of the \gls{UN}, respectively. Moreover, $\rho_{\textrm{UN}}[k]$ and $\alpha[k]$ denote the clock offset and clock skew of the \gls{UN} at time step $k$, correspondingly. Since the clock offsets of phase-locked \LOSANs are tracked in the \POSSYNCUKF, the state includes also a vector $\vec{\rho}_{\ell_{\textrm{AN}}}[k] = [\rho_{\ell_1}[k], \dots, \rho_{\ell_{L[k]}}[k]]\transpose$ consisting of the clock offsets of all $L[k]$ \glspl{AN} which are in \gls{LoS} condition with the \gls{UN} at time step $k$. However, when the \glspl{AN} are assumed to be synchronized among each other in the \POSCLOCKUKF, the \gls{AN} clock offset variables of the state can be neglected. It is to note that all the clock offsets are with respect to a chosen reference \gls{AN}.

In general, the \gls{CV} model is a widely used linear motion model in estimating the position of a moving object as done, e.g., in~\cite{werner_joint_2015}. The linear state transition matrix $\mat{F}$ as well as the constant covariance matrix $\mat{Q}$ of the state noise process for the CV model can be written as  \renewcommand{\arraystretch}{1.0}
\begin{align}
\mat{F} &= \blkdiag{ \left( 
\begin{bmatrix} \mat{I}_{3 \times 3} & \Delta t \mat{I}_{3 \times 3} \\ \mat{0}_{3 \times 3} & \mat{I}_{3 \times 3} \end{bmatrix}, 
%\mat{F}_{\textrm{AN}},
\begin{bmatrix}
1 & \Delta t \\
0 & 1\end{bmatrix}, \mat{I}_{L \times L} \right)},
\label{eq:state_transition2}
%\mat{F}_{\textrm{CA}} &= \blkdiag{ \left( 
%\begin{bmatrix} 
%\mat{I}_{3 \times 3} & \Delta t \mat{I}_{3 \times 3} & \Delta t^2 \mat{I}_{3 \times 3} \\ \mat{0}_{3 \times 3} & \mat{I}_{3 \times 3} & \Delta t \mat{I}_{3 \times 3} \\
%\mat{0}_{3 \times 3} & \mat{0}_{3 \times 3} & \mat{I}_{3 \times 3}
%\mat{F}_{\textrm{AN}} & \mat{A} \\
%\mat{0}_{3 \times 6} & \mat{I}_{3 \times 3}
%\end{bmatrix}, 
%\begin{bmatrix}
%1 & \Delta t \\
%0 & 1\end{bmatrix}, \mat{I}_{L \times L} \right)},
\end{align}
\renewcommand{\arraystretch}{1.4}
\begin{align}
\mat{Q} &= \blkdiag{ \left(\begin{bmatrix}
           \frac{\sigma_{v}^2 \Delta t^3 \mat{I}_{3\times3}}{3}   & \frac{\sigma_{v}^2 \Delta t^2 \mat{I}_{3 \times 3}}{2} \\
            \frac{\sigma_{v}^2 \Delta t^2\mat{I}_{3\times3}}{2}  & \sigma_{v}^2 \Delta t \mat{I}_{3\times3}\end{bmatrix}, \mat{Q}',\,\mat{Q}_{\rho} \right)},
%\mat{Q}_{\textrm{CA}} &= \blkdiag{ \left(\begin{bmatrix}
%           \frac{\sigma_{v}^2 \Delta t^3 \mat{I}_{3\times3}}{3}   & \frac{\sigma_{v}^2 \Delta t^2 \mat{I}_{3 \times 3}}{2} \\
%            \frac{\sigma_{v}^2 \Delta t^2\mat{I}_{3\times3}}{2}  & \sigma_{v}^2 \Delta t \mat{I}_{3\times3}\end{bmatrix}, \mat{Q}_{\textrm{Clk}}  \right)},
\end{align}
where the second and third arguments in $\mat{F}$ and $\mat{Q}$ correspond to the \gls{UN} and \LOSANs clock offset estimations, respectively, such that  
\begin{align}
&\mat{Q}' = \begin{bmatrix}
            \frac{\sigma_{\eta}^2 \Delta t^3}{3} & \frac{\sigma_{\eta}^2 \Delta t^2}{2}\\
            \frac{\sigma_{\eta}^2 \Delta t^2}{2} & \sigma_{\eta}^2 \Delta t
        \end{bmatrix},
&& \mat{Q}_{\rho}  = \sigma_{\rho}^2 \mat{I}_{L \times L},
%&& \mat{A} = \begin{bmatrix}
%            \Delta t^2 \mat{I}_{3 \times 3} \\ \Delta t \mat{I}_{3 \times 3}
%        \end{bmatrix},
\end{align}
where $\sigma_{\rho}^2$, $\sigma_v^2$, and $\sigma_{\eta}^2$ denote the variances of the \LOSANs clock offsets, \gls{UN} velocity and clock skew, respectively.

Moreover, the measurement function $\vec{h}_{\ell_i} : \mathbb{R}^{8+L} \rightarrow \mathbb{R}^3$ in~\eqref{eq:general_meas_model} that relates the state to the obtained \gls{DoA} and \gls{ToA} measurements $\vec{y}_{\ell_i}[k] = [\theta_{\ell_i}[k], \varphi_{\ell_i}[k], \tau_{\ell_i}[k]]\transpose$ estimated at the ${\ell_i}$th \gls{AN} can be written in the corresponding order as
\begin{align}
\vec{h}_{\ell_i}(\vec{s}[k]) =  \left[ \begin{array}{c}
                \arctan\left( \frac{\Delta y_{\ell_i}[k]}{\Delta x_{\ell_i}[k]}\right)\\
                \arctan\left( \frac{\Delta z_{\ell_i}[k]}{\| \vec{p}[k] - \vec{p}_{\ell_i} \|_{\textrm{2D}}}\right) \\
                \frac{\| \vec{p}[k] - \vec{p}_{\ell_i} \|_{\textrm{3D}}}{c} + (\rho_{{\ell_i}}[k]-\rho_{\textrm{UN}}[k])
                \end{array}\right],
                \label{eq:meas}
    %h_1 &= \arctan\left( \frac{\Delta y}{\Delta x}\right) \nonumber\\
    %h_2 &= \arctan\left( \frac{\Delta z}{\| x - p \|_{2D}}\right) \nonumber \\
    %h_3 &= \frac{\| x - p \|_{3D}}{c} + (\rho_{UN} - \rho_{AN}) \nonumber 
\end{align}
where $\Delta x_{\ell_i}[k],\,\Delta y_{\ell_i}[k],\,\Delta z_{\ell_i}[k]$ denote the distance between the \gls{UN} and the ${\ell_i}$th \gls{AN} in $x$-, $y$-, and $z$-directions, respectively. The distances between the \gls{UN} and the ${\ell_i}$th \gls{AN} in $xy$- and 3D-plane are denoted as $\| \vec{p}[k] - \vec{p}_{\ell_i} \|_{\textrm{2D}}$ and $\| \vec{p}[k] - \vec{p}_{\ell_i} \|_{\textrm{3D}}$, and finally $c$ represents the speed of light. Since it is assumed that all the clock offsets are with respect to a pre-defined reference \gls{AN}, the obtained \glspl{ToA} contain time offset differences between the \gls{UN} and received \LOSAN when the transmission is done in a time-stamping manner and, therefore, the difference $(\rho_{{\ell_i}}[k]- \rho_{\textrm{UN}}[k])$ needs to be added to the actual propagation delay in equation~\eqref{eq:meas}.

Finally, the measurement model functions and the corresponding measurements as well as the measurement noise terms for all \LOSANs need to be combined such that $\vec{h}(\vec{s}[k]) = [\vec{h}_{\ell_1}\transpose(\vec{s}[k]),\dots,\vec{h}_{\ell_L}\transpose(\vec{s}[k])]\transpose$, $\vec{y}[k] = [\vec{y}_{\ell_1}\transpose[k],\dots,\vec{y}_{\ell_L}\transpose[k]]\transpose$, and $\vec{w} \sim \mathcal{N}(0,\blkdiag{(\mat{R}_{k,{\ell_1}},\dots,\mat{R}_{k,{\ell_L}})})$. Thereafter, the models \eqref{eq:state_transition2} and \eqref{eq:meas} can be used in the proposed \DOA/\TOA Pos\&Clock and Pos\&Sync \glspl{UKF} as such, and after straightforward differentiation of the measurement function \eqref{eq:meas}, the models can be also applied to the extended 3D \DOA/\TOA Pos\&Clock and Pos\&Sync \glspl{EKF}.

%\begin{align}
%\mathbf{f}(\vec{s}) = \left[ \begin{array}{c}
%x + \frac{v_x}{\omega}\sin(\omega T) - \frac{v_y}{\omega}(1 - \cos(\omega T))\\
%y + \frac{v_x}{\omega}(1-\cos(\omega T)) + \frac{v_y}{\omega}\sin(\omega T)\\
%v_x \cos(\omega T) - v_y \sin (\omega T)\\
%v_x \sin (\omega T) + v_y \cos (\omega T)\\
%\omega\\
%\rho + T \alpha\\
%\alpha
%\end{array}\right] 
%\end{align}

\section{Numerical Evaluations and Analysis} \label{sec:results}

%In this section, the results obtained from the simulations and numerical evaluations are presented and analysed in a twofold manner. First, \gls{DoA} and \gls{ToA} estimation accuracies of both \gls{EKF} and \gls{UKF} based solutions are presented and discussed. Second, the positioning accuracies of the joint \gls{DoA}/\gls{ToA} Pos\&Clock/Sync \gls{UKF} and \gls{EKF} are compared with the positioning results obtained with the \gls{DoA}-only \gls{EKF} and \gls{UKF}.

In this section, numerical evaluations are carried out in order to demonstrate and evaluate the performance of the proposed methods in terms of \gls{DoA} and \gls{ToA} estimation, positioning, and clock offset estimation accuracy in the outdoor METIS Madrid map environment~\cite{metis_simulation_2013}. For the evaluations, the METIS map-based ray-tracing channel model is implemented using \gls{UTD} in order to model the propagation of received beacons~\cite{metis_channel_2015} as realistically as possible. Furthermore, the transmit power of the tracked \glspl{UN} is set to \SI{10}{dBm}, and interfering \glspl{UN} with the same transmit power are placed on the map randomly \SI{250}{m} away from the \gls{UN} with a density of \SI{1000}{interferers/km^2}.

The considered \gls{5G} network is assumed to deploy \gls{OFDMA}-based radio access with \SI{240}{kHz} subcarrier spacing and \SI{5}{MHz} beacon bandwidth, for one \gls{UN}, comprising of $20$ pilot subcarriers~\cite{kela_borderless_2015}. In addition, subframes of length \SI{0.2}{ms} containing $14$ \gls{OFDM} symbols are incorporated into the radio frame structure. Furthermore, beacons of the \glspl{UN} within a specific \gls{AN} coordination area are assumed to be orthogonal through proper time and frequency multiplexing. Moreover, the proposed filters are updated only every \SI{100}{ms} to facilitate communications between the \glspl{AN} and the central unit of the network as well as relax the \gls{UN}'s energy requirements, while measurements from only $L = 2$ closest \LOSANs are fused for positioning purposes.

%\begin{figure}[!t]
%    \centering
%    \includegraphics[width=3.5in]{figures/3d_pos_errors_final}
%    \caption{}
%    \label{fig:3d_pos_errors}
%\end{figure}

In the evaluations, \DOA and \TOA estimation as well as positioning and synchronization performance is analysed by averaging over multiple 3D \gls{UN} test trajectories. One half of the random trajectories demonstrates a usual vehicle movement in an urban environment with a constant vertical position whereas the other half of the trajectories contains more variation in a vertical direction illustrating applications such as drones. Motion model of the vehicles follows the empirical polynomial model in~\cite{akcelik_acceleration_1987} with a maximum speed of \SI{50}{km/h}, and the same model is also adopted for drone trajectories containing landings and take-offs as well as short halts on the ground after landing.

%Small \SI{5}{MHz} carrier bandwidth for continuously allocated UL pilots (More users can be multiplexed to the channel simultaneously). Subcarrier spacing is chosen to be \SI{240}{kHz} which equals to $2^4$ times commonly envisioned \SI{15}{kHz} subcarrier spacing. \gls{UTD}

\begin{figure}[!t]
    \centering
    \includegraphics[width=3.5in, trim=0cm 1.2cm 0cm 0.8cm, clip]{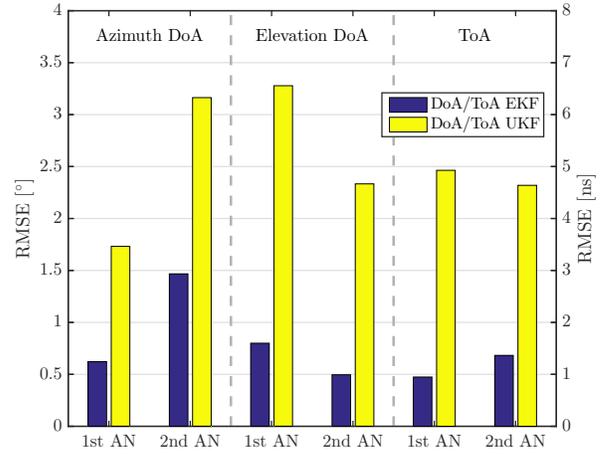}
    \caption{\gls{DoA} and \gls{ToA} estimation results for \gls{DoA}/\gls{ToA} UKF based processing at two closest \gls{LoS}-\glspl{AN} in comparison with the extended \gls{EKF}-based approach.\\\mbox{}\\}
    \label{fig:doa_toa_results}
    \vspace{\spaceunderfig}
    \vspace{4.2pt}
\end{figure}

\subsection{DoA and ToA Estimation}

In order to analyse the estimation accuracy of the proposed \DOATOAUKF in comparison with the \DOATOAEKF, \glspl{RMSE} of \DOA and \TOA tracking for the two closest \LOSANs are depicted in Fig.~\ref{fig:doa_toa_results}, after averaging over multiple random trajectories on the Madrid grid. Yellow bars in Fig.~\ref{fig:doa_toa_results} represent \glspl{RMSE} for the \DOATOAUKF whereas blue bars illustrate the performance of the \DOATOAEKF.

In general, the obtained results are extremely accurate for both methods. However, the \DOATOAEKF seems to outperform the proposed \gls{UKF}-based method in both \DOA and \TOA estimation. Based on the observations, we noticed that the \DOATOAUKF experienced some disadvantageous divergence with the mapped sigma points and, therefore, the algorithm parameter $\lambda$ was set to a large value. Although the results are relatively accurate with the \DOATOAUKF as well, the filter may not fully achieve all of its potential due to the sigma point mapping thus raising a possible topic for the future work. In particular, the sigma-points should exploit the manifold in which the angular parameters are defined. Currently, the conventional form of the \gls{UKF} and similarly for the sigma-point generation have been employed.

\subsection{Positioning, Clock and Network Synchronization}

\begin{figure}[!t]
    \centering
    \includegraphics[width=3.5in, trim=0cm 1.2cm 0cm 0.8cm, clip]{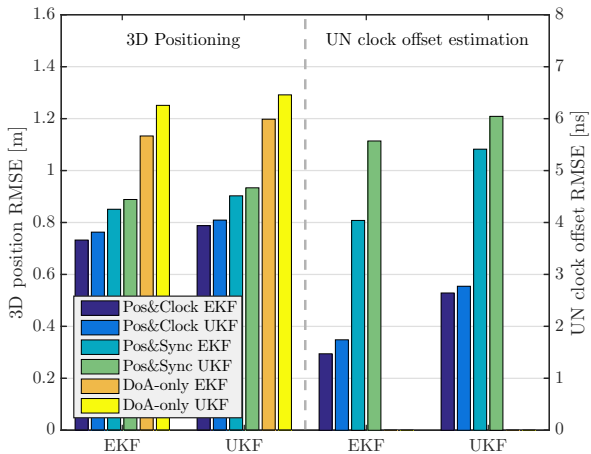}
    \caption{3D positioning and \gls{UN} clock offset estimation RMSEs for all estimation methods and for both \DOATOAEKF and \DOATOAUKF scenarios, denoted as EKF and UKF, respectively, after averaging over multiple random \gls{UN} trajectories.}
    \label{fig:clk_errors}
    \vspace{\spaceunderfig}
\end{figure}

In the beginning of evaluations, states of the proposed positioning and synchronization methods are initialized. The initial position of the \gls{UN} is determined using the known positions of the initial \LOSANs in the \gls{CL} method. Variance of the initial position, denoted as $\sigma_{p,0}^2$ is set to a large value using the distance between the initial \gls{UN} estimate and \LOSANs. Furthermore, the initial velocity is set to $\vec{v}[0] \sim \mathcal{N}(0,(\SI{5}{m/s})^2\mat{I}_{3 \times 3})$ since no additional information in the initialization is used. In the beginning, the clock offset and skew of the \glspl{UN} are set according to $\rho_{\textrm{UN}}[0] \sim \mathcal{N}(0,(\SI{100}{\micro s})^2)$ and $\alpha[0] \sim \mathcal{N}(\SI{25}{ppm},(\SI{30}{ppm})^2)$, respectively, based on~\cite{kim_tracking_2012}. In the phase-locked \glspl{AN} scenario, the clock offsets of the \glspl{AN} are initialized similarly to the \glspl{UN}. In addition, the standard deviation of the clock skew process noise is set to $\sigma_{\eta} = 6.3\cdot10^{-8}$ according to~\cite{kim_tracking_2012} whereas the same value within the filters is increased to $\sigma_{\eta} = 10^{-4}$ leading to a much better overall performance. Furthermore, the standard deviation of the velocity within the \gls{UKF} and \gls{EKF} is set to $\sigma_v = \SI{3.5}{m/s}$. Due to the best overall performance, the algorithm parameters for the both \DOA/\TOA Pos\&Clock and Pos\&Sync \glspl{UKF} are set such that $\alpha = 10^{-3}$, $\beta = 2$, and $\kappa = 0$.

The performance of the proposed \DOATOAUKF and extended \DOATOAEKF is evaluated by tracking \glspl{UN} through random trajectories on the Madrid map. The presented methods in both network synchronization scenarios are carried out using \DOA and \TOA estimates from both \DOATOAUKF and \DOATOAEKF, and finally the results are compared with the \DOA-only \gls{EKF} and \gls{UKF}. The achieved 3D positioning and \gls{UN} clock offset estimation RMSE results for each filtering method are illustrated in Fig.~\ref{fig:clk_errors}. Based on the observations, the \LOSANs clock offset errors are almost identical to the clock offset errors of the \glspl{UN} and hence these errors are not visualized in this paper. In addition, 2D positioning as well as vertical positioning RMSEs are shown separately in Table~\ref{tab:results}.

Based on the results, the overall positioning performance of the proposed \gls{UKF}- and \gls{EKF}-based methods is extremely high. As shown in Fig.~\ref{fig:clk_errors}, the envisioned sub-meter positioning accuracy in \gls{5G}~\cite{5g-ppp_2015} can be achieved even in 3D scenarios when accurate \TOA measurements are available. Furthermore, as expected, the positioning accuracy of the proposed methods is more accurate in the case of synchronized \glspl{AN}, i.e., the \DOA/\TOA Pos\&Clock filters, compared to the phase-locked \glspl{AN}, i.e., the \DOA/\TOA Pos\&Sync filters. In general, a clear relation between the \DOA and \TOA estimation accuracy, and positioning and synchronization accuracy can also be seen by comparing the results in Fig.~\ref{fig:doa_toa_results} and Fig.~\ref{fig:clk_errors}. The overall positioning as well as synchronization results are both better when the \DOATOAEKF was used due to more accurate \DOA and \TOA estimation. Despite the fact that 2D positioning results are better for the \DOATOAUKF than \DOATOAEKF, a clear decrease in vertical positioning can be noticed in Table~\ref{tab:results}. Although the \gls{EKF}-based measurement estimation method is slightly outperforming the proposed \gls{UKF}-based solution, there is no significant difference between the corresponding positioning and synchronization methods. Furthermore, the obtained \gls{UN} clock offset estimates, depicted in Fig.~\ref{fig:clk_errors}, illustrate also the synchronization capability of the proposed methods. In the case of synchronized network, the \gls{UN} clock offset estimation accuracy is even below \SI{2}{ns} at its best. However, the very appealing below \SI{10}{ns} clock offset estimates for the \glspl{UN} as well as \LOSANs are also achieved with the phase-locked \glspl{AN}.

The behavior of the proposed positioning methods is illustrated in the videos at \mbox{\texttt{\url{http://www.tut.fi/5G/GLOBECOM16}}}.

\renewcommand{\arraystretch}{1.0}
\begin{table}[!t]
    \centering
        \caption{Partitioned 2D and vertical positioning RMSEs for the both \DOATOAEKF and \DOATOAUKF measurement estimation scenarios.}
    \label{tab:results}
    \small
\begin{tabular}{l|c c|c c}
&\multicolumn{2}{c|}{\DOATOAEKF} & \multicolumn{2}{c}{\DOATOAUKF}\\
& 2D (m)& z (m) & 2D (m) & z (m)\\
\hline\hline
Pos\&Clock EKF & 0.71 & 0.19 & 0.70 & 0.32 \\
Pos\&Clock UKF & 0.73 & 0.21 & 0.72 & 0.33\\
Pos\&Sync EKF & 0.83 & 0.20 & 0.82 & 0.34\\
Pos\&Sync UKF & 0.86 & 0.23 & 0.84 & 0.35\\
DoA-only EKF & 1.11 & 0.22 & 1.12 & 0.35\\
DoA-only UKF & 1.22 & 0.27 & 1.20 & 0.40\\
\hline \hline
\end{tabular}
    \vspace{-5pt}
\end{table}
 
%In this paper, two different scenarios for synchronization within a network are assumed. In the first scenario, \glspl{UN} are assumed to have unsynchronized clocks when necessary timing and frequency synchronization to avoid \gls{ICI} and \gls{ISI} is achieved. On the other hand, \glspl{AN} in such scenario are assumed to have synchronized clocks among each other. In the second and more realistic scenario, clocks within \glspl{AN} are set to phase-locked, i.e., the clock offsets of \glspl{AN} are not fundamentally varying over the real time, whereas clocks within \glspl{UN} are assumed to be unsynchronized. Aforementioned synchronized and phase-locked clocks within \glspl{AN} can be obtained using a time reference from, e.g., GPS, or by communicating a reference time from a central unit of the network to the \glspl{AN}. However, these methods surely increase the signaling overhead. 

%\begin{figure}[!t]
%    \centering
%    \includegraphics[scale=0.5]{figures/UKF-EKF_pos_comparison}
%    \caption{Positioning performance of Sync\&Pos EKF and Sync\&Pos UKF with \SI{4.8}{MHz} and \SI{9.6}{MHz} \gls{RS} bandwidths, and with two different motion models.}
%    \label{fig:pos_results}
%\end{figure}

%\begin{figure}[!t]
%    \centering
%    \includegraphics[scale=0.5]{figures/UKF-EKF_clk_comparison}
%    \caption{\gls{UN} offset estimation performance of Sync\&Pos EKF and Sync\&Pos UKF with \SI{4.8}{MHz} and \SI{9.6}{MHz} \gls{RS} bandwidths, and with two different motion models.}
%    \label{fig:pos_results}
%\end{figure}
\section{Conclusion} \label{sec:conclusion}

In this paper, a \gls{UKF}-based solution for 3D positioning and network synchronization in future \gls{5G} \glspl{UDN} was proposed. First, \glspl{UKF} are deployed for estimating and tracking the \DOAs and \TOAs of devices at \glspl{AN} using \gls{UL} beacons. Thereafter, a novel \gls{UKF} solution was proposed to fuse the obtained \DOAs and \TOAs from \LOSANs into 3D device position and clock offset estimates while considering realistic clock offsets between devices and \glspl{AN} as well as mutual clock offsets among the \glspl{AN}. %Thus, the proposed solution can provide not only a highly accurate 3D position estimates but also valuable clock offset estimates as a by-product. 
Furthermore, a similar \gls{EKF}-based 2D positioning approach was extended for comparison purposes to cover also 3D positioning scenarios. Finally, comprehensive numerical evaluations were carried out considering different 3D motion scenarios for vehicles and \glspl{UAV} in the realistic Madrid grid environment together with the METIS map-based ray-tracing channel model. Based on the obtained results, below one meter 3D positioning and tracking accuracy can be achieved using the proposed \gls{EKF}- and \gls{UKF}-based methods. Simultaneously, the proposed methods also provide extremely accurate nanosecond-scale clock offset estimates for unsynchronized clocks as a valuable by-product.

\linespread{0.87}
\bibliographystyle{IEEEtran}
\bibliography{main}

\end{document}